\documentclass[epj]{svjour}

\usepackage{graphicx}
\usepackage{fancyhdr}
\def\makeheadbox{{
 }}
\def\mytitle{My title} 
\def\myauthors{My name}  
\def\mytype{My type of session}
\def\mysession{My session}
\def\mytitle{The lightest neutralino in the MNSSM} %Put your title here!
\def\myauthors{R.~Nevzorov}    %Put your name here!
\def\mytype{Contributed Talk}    
\def\mysession{Cosmology and Astrophysics}
\begin{document}
\title{The lightest neutralino in the MNSSM}
\author{S.~Hesselbach\inst{1}
\and
D.~J.~Miller\inst{2}% etc
\and
G.~Moortgat-Pick\inst{3}% etc
\and
R.~Nevzorov\inst{2}% etc
\thanks{\emph{Email:} r.nevzorov@physics.gla.ac.uk}
\and
M.~Trusov\inst{4}% etc
}                     % Do not remove
\institute{
School of Physics and Astronomy, University of Southampton, Southampton, SO17 1BJ, U.K.
\and
Department of Physics and Astronomy, University of Glasgow, Glasgow, G12 8QQ, U.K.
\and
IPPP, University of Durham, Durham, DH1 3LE, U.K.
\and 
Theory Department, ITEP, Moscow, Russia}
\date{}
\abstract{We examine the allowed mass range of the lightest neutralino 
within the Minimal Non--minimal Supersymmetric Standard Model.
Being absolutely stable if R-parity is conserved this
lightest neutralino is a candidate for the dark matter of the universe.
We establish the theoretical upper bound on the  
lightest neutralino mass and obtain an approximate solution for this
mass.
 % end of PACS codes
} %end of abstract

\maketitle
\section{Introduction}
\label{intro}

The existence of dark matter is now a well established fact.
Recent astrophysical and cosmological observations indicate that 
the Universe contains approximately five times more exotic matter 
than ordinary matter. It corresponds to 22\%-25\% of the energy density 
of the Universe \cite{Bertone:2004pz}. This exotic matter exists in the form 
of non--baryonic, non--luminos (dark) matter. Although the microscopic 
composition of dark matter remains a mystery it is clear that it can not 
consist of any elementary particles which have been discovered so far. 

The\, minimal\, supersymmetric\, (SUSY)\, standard\, model (MSSM) 
which is the best motivated extension of the SM, provides
a good candidate for the cold dark matter component of the Universe. 
If R--parity is conserved, the lightest supersymmetric particle (LSP) in the 
MSSM is absolutely stable and can play the role of dark matter. In most 
supersymmetric scenarios the LSP is the lightest neutralino. 
Neutralinos naturally provide the correct relic abundance of dark matter if 
these particles have masses of a few hundred GeV. Furthermore in this 
case they behave as cold (non--relativistic) dark matter which explains well 
the large scale structure of the Universe.

In\, spite\, of\, its\, attractiveness\, the MSSM\, has\, some\, unattractive features as well.
One of them is the $\mu$--problem. The MSSM superpotential contains only one 
bilinear term $\mu (\hat{H}_1 \epsilon  \hat{H}_2)$ which is present before 
supersymmetry is broken. In order to get the correct pattern of electroweak
symmetry breaking, the parameter $\mu$ is required to be of the order of 
the electroweak scale. While the corresponding coupling is stable under 
quantum corrections as a result of supersymmetry, it is rather difficult 
to explain within Grand Unified Theories (GUTs) or supergravity (SUGRA) 
why the dimensionful quantity should be so much smaller than the Planck 
or Grand Unification scale.

An elegant solution of the $\mu$--problem naturally appears in the framework
of the Next--to--Minimal Supersymmetric Standard Model (NMSSM) in which a
$Z_3$ symmetry forbids any bilinear terms in the superpotential but allows 
the interaction of an extra singlet field $S$ with the Higgs doublets 
$H_u$ and $H_d$: $\lambda \hat{S}(\hat{H}_d\epsilon\hat{H}_u)$. 
At the electroweak (EW) scale the superfield $\hat{S}$ gets non-zero vacuum 
expectation value ($\langle S \rangle =s/\sqrt{2}$) and an effective $\mu$-term 
($\mu=\lambda s/\sqrt{2}$) of the required size is automatically generated. 
But the NMSSM itself is not without problems. The vacuum expectation values of 
the Higgs fields break the $Z_3$ symmetry. This leads to the formation of domain 
walls in the early Universe between regions which were causally disconnected 
during the period of electroweak symmetry breaking  \cite{okun}. Such domain structures 
of the vacuum create unacceptably large anisotropies in the cosmic microwave background 
radiation  \cite{vilenkin}. In an attempt to break the $Z_3$ symmetry, operators that are suppressed 
by powers of the Planck scale could be introduced, but these operators give\, rise\, 
to\, quadratically\, divergent\, tadpole\, contributions,\, which\, destabilise the mass 
hierarchy  \cite{abel}. The dangerous operators can be eliminated if an invariance under 
$Z_2^R$ or $Z_{5}^R$ symmetries is imposed  \cite{Panagiotakopoulos:1998yw}--\cite{Panagiotakopoulos:1999ah}. 
A linear term $\Lambda\hat{S}$ in the superpotential is induced 
by high order operators. It is too small to affect the mass hierarchy but large enough 
to prevent the appearance of domain walls. The superpotential of the corresponding 
simplest extension of the MSSM --- the Minimal Non--minimal Supersymmetric Standard Model 
(MNSSM) can be written as   \cite{Panagiotakopoulos:1999ah}--\cite{Dedes:2000jp}
\begin{equation}
W_{MNSSM}=\lambda \hat{S}(H_d \epsilon H_u)+\xi \hat{S}
+W_{MSSM}(\mu=0)\,.
\label{2}
\end{equation}

\section{Theoretical restrictions on the lightest neutralino mass}
\label{sec:2}

The neutralino sector in SUSY models is formed by the superpartners of neutral gauge and Higgs bosons.
Since the sector responsible for elecroweak symmetry breaking in the
MNSSM contains an extra singlet field, the
neutralino sector of this model includes one extra component besides the four MSSM ones.
This is an additional Higgsino $\tilde{S}$ (singlino) which is the fermion component of the SM singlet
superfield $S$. In the field basis $(\tilde{B},\,\tilde{W}_3,\,\tilde{H}^0_d,\,\tilde{H}^0_u,\,\tilde{S})$
the neutralino mass matrix reads
{\small
\begin{equation}
\begin{array}{l}
M_{\tilde{\chi}^0}=\\
\left(
\begin{array}{ccccc}
M_1                  & 0                  & -M_Z s_W c_{\beta}   & M_Z s_W s_{\beta}  & 0 \\[2mm]
0                    & M_2                & M_Z c_W c_{\beta}    & -M_Z c_W s_{\beta} & 0 \\[2mm]
-M_Z s_W c_{\beta}   & M_Z c_W c_{\beta}  &  0                   & -\mu_{eff}         & -\displaystyle\frac{\lambda v}{\sqrt{2}} s_{\beta} \\[2mm]
M_Z s_W s_{\beta}    & -M_Z c_W s_{\beta} & -\mu_{eff}           &  0                 & -\displaystyle\frac{\lambda v}{\sqrt{2}} c_{\beta} \\[2mm]
0                    & 0                  & -\displaystyle\frac{\lambda v}{\sqrt{2}} s_{\beta}  & -\displaystyle\frac{\lambda v}{\sqrt{2}} c_{\beta} &
0
\end{array}
\right),
\end{array}
\label{1}
\end{equation}
}
\noindent
where $M_1$ and $M_2$ are $U(1)_Y$ and $SU(2)$ gaugino masses while $s_W=\sin\theta_W$, $c_W=\cos\theta_W$, $s_{\beta}=\sin\beta$, 
$c_{\beta}=\cos\beta$ and $\mu_{eff}=\lambda s/\sqrt{2}$. Here we introduce $\tan\beta=v_2/v_1$ and 
$v=\sqrt{v_1^2+v_2^2}=246\,\mbox{GeV}$, where $s$, $v_1$ and $v_2$ are
the vacuum expectation values of the singlet field $S$ and 
of the Higgs doublets fields $H_d$ and $H_u$, respectively.

The top--left $4\times 4$ block of the mass matrix (\ref{1}) contains the neutralino mass matrix of the MSSM where 
the parameter $\mu$ is replaced by $\mu_{eff}$. From Eq.(\ref{1}) one can easily see that the neutralino spectrum 
in the MNSSM may be parametrised in terms of
\begin{equation}
\lambda\,,\qquad \mu_{eff}\,,\qquad \tan\beta\,, \qquad M_1\,,\qquad M_2\,.
\label{3}
\end{equation}
In\, supergravity\, models\, with\, uniform\, gaugino\, masses\, at\, the\, Grand\, Unification\, scale\, the\, renormalisation\, 
group flow yields a relationship between $M_1$ and $M_2$ at the EW scale, i.e. $ M_1\simeq 0.5 M_2$.
The chargino masses in SUSY models are defined by the mass parameters $M_2$ and $\mu_{eff}$. 
LEP searches for SUSY particles set a lower limit on the chargino mass of about $100\,\mbox{GeV}$. This\, lower\, bound\,
constrains\, the\, parameter\, space\, of\, the\, MNSSM\, restricting\, the\, absolute\, values\, of\, the\, effective\, 
$\mu$-term\, and\, $M_2$\, from\, below,\, i.e. $|M_2|$, $|\mu_{eff}|\ge 90-100\,\mbox{GeV}$.

In\, general\, the\, eigenvalues\, of\, the neutralino\, mass\, matrix\, can\, be\, complex.\, This\, prevents\, the\, 
establishing\, of\, any\, theoretical\, restrictions\, on\, the\, masses\, of\, neutralinos.\, In\, order\, to\, 
find\, appropriate\, bounds\, on\, the\, neutralino\, masses\, \cite{neutralino}\, it\, is\, much\,
more\, convenient\, to\, consider\, the matrix\, $M_{\tilde{\chi}^0} M^{\dagger}_{\tilde{\chi}^0}$\, whose\, eigenvalues\, are\,
positive\, definite\, and\, equal\, to\, the\, absolute\, values\, of\, the\, neutralino\, mass\, squared.\, In\, the\, basis\, 
$\left(\tilde{B},\,\tilde{W}_3,\,-\tilde{H}^0_d s_{\beta}+\tilde{H}^0_u c_{\beta},\, \tilde{H}^0_d c_{\beta}+\tilde{H}^0_u s_{\beta},
\,\tilde{S}\right)$ the bottom-right $2\times 2$ block of $M_{\tilde{\chi}^0} M^{\dagger}_{\tilde{\chi}^0}$ takes the form
\begin{equation}
\left(
\begin{array}{cc}
|\mu_{eff}|^2+\sigma^2 \qquad\qquad           & \nu^{*}\mu_{eff} \\
 \nu\mu^{*}_{eff}                             & |\nu|^2
\end{array}
\right),
\label{4}
\end{equation}
where $\sigma^2=M_Z^2\cos^2 2\beta+|\nu|^2\sin^2 2\beta,\,\nu=\lambda v/\sqrt{2}$.
Since the minimal eigenvalue of any hermitian matrix is less than its smallest diagonal element
at least one neutralino in the MNSSM is always light. Indeed, in the
considered case the mass interval
of the lightest neutralino is limited from above by the bottom--right diagonal
entry of matrix (\ref{4}), i.e. $|m_{\chi^0_1}|\le |\nu|$. Therefore
in contrast with the MSSM
the lightest neutralino in the MNSSM remains light even when the SUSY breaking scale tends to infinity.

The obtained theoretical bound on the lightest neutralino mass can even be improved significantly.
Since we can always choose the field basis in such a way that this $2\times 2$ submatrix of 
$M_{\tilde{\chi}^0} M^{\dagger}_{\tilde{\chi}^0}$ becomes diagonal its eigenvalues also restrict 
the mass interval of the lightest neutralino. In particular, the absolute value of 
the lightest neutralino mass squared has to be always less than or
equal to the minimal eigenvalue $\mu^2_0$ of 
the corresponding submatrix, i.e.
\begin{equation}
\begin{array}{c}
|m_{\chi^0_1}|^2\le \mu_{0}^2=\displaystyle\frac{1}{2}\biggl[|\mu_{eff}|^2+\tilde{\sigma}^2+|\nu|^2\quad\qquad\\[1mm]
\quad\qquad-\sqrt{\biggl(|\mu_{eff}|^2+\tilde{\sigma}^2+|\nu|^2\biggr)^2-4|\nu|^2\tilde{\sigma}^2}\biggr]\,.
\end{array}
\label{5}
\end{equation}
The value of $\mu_0$ reduces with increasing $|\mu_{eff}|$. It reaches its maximum value, i.e.
$\mu_0^2=\min\biggl\{\tilde{\sigma}^2,\,|\nu|^2 \biggr\}$, when $\mu_{eff}\to 0$. Taking into account
the restriction on the effective $\mu$--term coming from LEP searches and the theoretical upper bound
on the Yukawa coupling $\lambda$ which is caused by the requirement of the validity of perturbation theory up to
the Grand Unification scale ($\lambda<0.7$) we find that $\mu_0^2<0.8 M_Z^2$.
When $|m_{\chi^0_1}|$ is close to its maximum value the lightest neutralino mass is predominantly a superposition
of $U(1)_Y$ gaugino and singlino.

Here it is worth to notice that at large values of the term $\mu_{eff}$ the theoretical restriction on
$|m_{\chi^0_1}|$ (\ref{5}) tends to zero independently of the value of
$\lambda$. Indeed, for $|\mu_{eff}|^2\gg M_Z^2$ 
we have
\begin{equation}
|m_{\chi^0_1}|^2\le \displaystyle\frac{|\nu|^2\tilde{\sigma}^2}{\biggl(|\mu_{eff}|^2+\tilde{\sigma}^2+|\nu|^2\biggr)}\,.
\label{9}
\end{equation}
Thus in the considered limit the lightest neutralino mass is significantly smaller than $M_Z$ even for the
appreciable values of $\lambda$ at tree level.

\section{Approximate solution}
\label{sec:3}

The masses of the lightest neutralino can be computed numerically by solving the characteristic
equation $\mbox{det}\left(M_{\tilde{\chi}^0}-\kappa I\right)=0$. In the MNSSM the corresponding characteristic polynomial
has degree $5$ because the neutralino spectrum is described by a $5\times 5$ mass matrix. After a few simple algebraic
transformations we get
\begin{equation}
\begin{array}{c}
\mbox{det}\left(M_{\tilde{\chi}^0}-\kappa I\right)=
\biggl(M_1M_2-(M_1+M_2)\kappa+\kappa^2\biggr)\times\\[3mm]
\times\biggl(\kappa^3-(\mu_{eff}^2+\nu^2)\kappa+\nu^2\mu_{eff}\sin 2\beta\biggr)+
\end{array}
\label{10}
\end{equation}
$$
+M_Z^2\biggl(\tilde{M}-\kappa\biggr)\biggl(\kappa^2+\mu_{eff}\sin 2\beta \kappa-\nu^2\biggr)=0\,,
$$
where $\tilde{M}=M_1 c_W^2 + M_2 s^2_W$. Although one can find a numerical solution of Eq.~(\ref{10}) for each set of
the parameters (\ref{3}) it is rather interesting to explore analytically the dependence of the lightest neutralino mass on these parameters.
In order to perform such an analysis it is worthwhile to derive at least an approximate solution of the characteristic
equation (\ref{10}). Such an approximate solution can be obtained in the limit when one of the eigenvalues of the mass 
matrix (\ref{1}) goes to zero. Indeed, if $\kappa\to 0$ we can ignore all higher order terms with respect to $\kappa$ in
the characteristic equation keeping only terms which are proportional to
$\kappa$ and $\kappa^2$ as well as the $\kappa$--independent ones.
In that case, Eq.~(\ref{10}) takes the form
\begin{equation}
\alpha\kappa^2-\beta\kappa+\gamma=0,
\label{11}
\end{equation}
where
\begin{equation}
\begin{array}{rcl}
\alpha&=&1+\displaystyle\frac{\nu^2-M_Z^2}{\mu_{eff}^2+\nu^2}\frac{\mu_{eff}\sin 2\beta}{M_1+M_2}+\qquad\qquad\qquad\\[1mm]
&&\qquad\qquad\qquad\qquad+\displaystyle\frac{M_Z^2}{\mu_{eff}^2+\nu^2}\frac{\tilde{M}}{M_1+M_2}\,,
\end{array}
\label{12}
\end{equation}
\begin{equation}
\begin{array}{rcl}
\beta&=&\displaystyle\frac{M_1 M_2}{M_1+M_2}+\biggl(\frac{\nu^2}{\mu_{eff}^2+\nu^2}-
\displaystyle\frac{M_Z^2}{\mu_{eff}^2+\nu^2}\frac{\tilde{M}}{M_1+M_2}\biggr)\times
\end{array}
\label{13}
\end{equation}
$$
\qquad\qquad\times \mu_{eff}\sin 2\beta-\displaystyle\frac{M_Z^2\nu^2}{(M_1+M_2)(\mu_{eff}^2+\nu^2)}\,,
$$
\begin{equation}
\begin{array}{rcl}
\gamma&=&\displaystyle\frac{\nu^2}{\mu_{eff}^2+\nu^2}\biggl(\frac{M_1M_2}{M_1+M_2}\mu_{eff}\sin 2\beta\qquad\qquad\\[1mm]
&&\qquad\qquad\qquad\qquad\qquad-\displaystyle\frac{\tilde{M}}{M_1+M_2}M_Z^2\biggr)\,.
\end{array}
\label{14}
\end{equation}
In the MNSSM there is a good justification for applying this
method. As we argued in the previous section, the mass of the lightest neutralino is limited from above and
an upper bound on $|m_{\chi^0_1}|$ tends to be zero with raising of $|\mu_{eff}|$ or decreasing of $\lambda$.

One can simplify the reduced form of the characteristic equation (\ref{11}) even further taking into account
that the second and last terms in the Eq.~(\ref{12}) can be neglected since they are much smaller than unity in most of
the phenomenologically allowed region of the MNSSM parameter space. Then the mass of the lightest neutralino can be approximated by
\begin{equation}
|m_{\chi^0_1}|=\mbox{Min}\left\{\frac{1}{2}\biggl|\beta\pm\sqrt{\beta^2-4\gamma}\biggr|\right\}\,.
\label{15}
\end{equation}

\begin{figure}
\hspace{0cm}{$|m_{\chi_1^0}|$}\\
\includegraphics[width=0.45\textwidth, keepaspectratio=true
%height=0.15\textwidth,angle=0
]{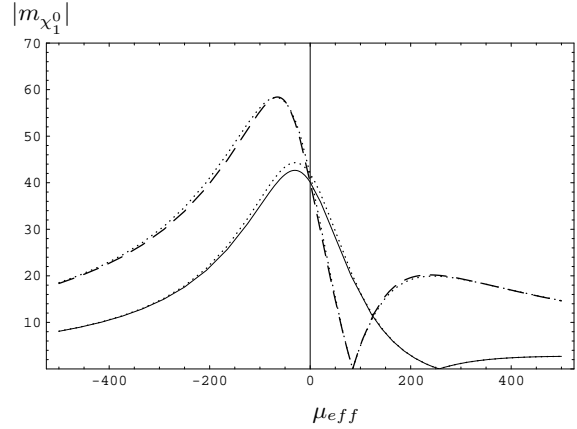}
\hspace*{4cm}{$\mu_{eff}$}
\caption{Lightest neutralino mass versus $\mu_{eff}$ in the MNSSM. Solid and dashed lines correspond to $\tan\beta=10$ 
and $3$ respectively. Dotted lines represent the approximate solution for the lightest neutralino mass.
Other parameters are fixed as follows $\lambda=0.7$, $M_1=0.5 M_2$, $M_2=200\,\mbox{GeV}$.}
\label{fig:1}      
\end{figure}

In Figs.1--3 we plot both the numerical and the approximate solutions for the lightest neutralino mass as a function of $\mu_{eff}$, $M_2$
and $\tan\beta$. In the present study we assume that all parameters
(\ref{3}) appearing in the neutralino mass matrix are real.
We also choose $M_1=0.5 M_2$ and $\lambda=0.7$ which is the largest possible value of $\lambda$ that does not spoil the
validity of perturbation theory up to the GUT scale. Figs.1--3 demonstrate that the approximate solution (\ref{15}) describes
the numerical one with relatively high accuracy even for $M_2\simeq\mu_{eff}\simeq 150\,\mbox{GeV}$, see Fig.3\,. One can also see that
the mass of the lightest neutralino may be very small or even takes zero value for appreciable values of $\lambda$.
This happens because the determinant of the neutralino mass matrix (\ref{1}) tends to zero for a certain relation between
the parameters
\begin{equation}
M_1 M_2 \mu_{eff}\sin 2\beta=\tilde{M} M_Z^2\,.
\label{16}
\end{equation}
It is worth noticing that condition (\ref{16}) is fulfilled
automatically when $M_1\sim M_2\to 0$. Thus the absolute
value of the lightest neutralino mass vanishes only once in Figs.1 and 3 and twice in Fig.2\,. 

\begin{figure}
\hspace{0cm}{$|m_{\chi_1^0}|$}\\
\includegraphics[width=0.45\textwidth, keepaspectratio=true
%height=0.15\textwidth,angle=0
]{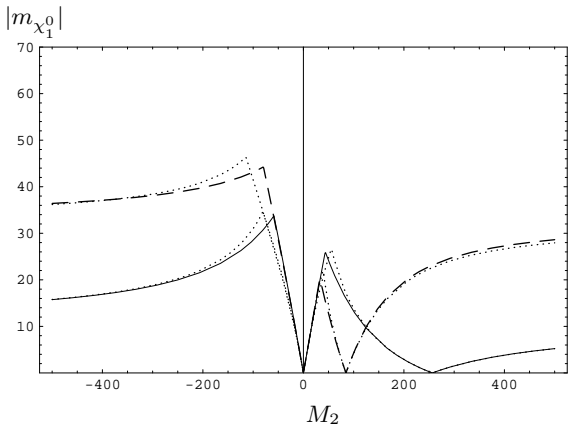}
\hspace*{4cm}{$M_{2}$}
\caption{The mass of the lightest neutralino in the MNSSM as a function of $M_2$ for 
$\lambda=0.7$, $M_1=0.5 M_2$, $\mu_{eff}=200\,\mbox{GeV}$. The notations are the same as 
in Fig.~1.}
\label{fig:2}       
\end{figure}

At large $|\mu_{eff}|,\, |M_2|, |M_1|\gg M_Z$ the value of $|m_{\chi_1^0}|$ decreases with raising 
of the absolute values of the effective $\mu$--term and soft gaugino masses (see Figs.1--2).
From Fig.1--3 it becomes clear that the difference between the numerical and approximate solutions 
reduces when $|\mu_{eff}|$, $|M_1|$ and $|M_2|$ grow. If either $|\mu_{eff}|$ or $|M_1|$ and $|M_2|$ 
are much larger than $M_Z$, $\beta^2\gg \gamma$, the approximate solution for the 
lightest neutralino mass can be presented in a more simple form:
\begin{equation}
|m_{\chi^0_1}|\simeq\biggl|\frac{\gamma}{\beta}\biggr|\simeq\displaystyle\frac{|\mu_{eff}|\nu^2\sin 2\beta}{\mu^2_{eff}+\nu^2}\,.
\label{17}
\end{equation}
According to Eq.(\ref{17}) the mass of the lightest neutralino is inversely proportional to the term $\mu_{eff}$.
It vanishes when $\lambda$ tends to zero. In the limit $\lambda\to 0$ the equations for the extrema of the Higgs effective
potential that determine the position of the physical vacuum imply
that the vacuum expectation value of the singlet field rises
as $M_Z/\lambda$. In other words the correct breakdown of electroweak symmetry breaking requires $\mu_{eff}$ to remain constant
when $\lambda$ goes to zero. Therefore, it follows from Eq.~(\ref{17}) that the mass of the lightest neutralino is proportional
to $\lambda^2$ at small values of $\lambda$. At this point the approximate solution (\ref{17}) improves the theoretical
restriction on the lightest neutralino mass derived in the previous section because for small values of $\lambda$
the upper bound (\ref{9}) implies that $|m_{\chi^0_1}|$ is proportional to $\lambda$.

\begin{figure}
\hspace{0cm}{$|m_{\chi_1^0}|$}\\
\includegraphics[width=0.45\textwidth, keepaspectratio=true
%height=0.15\textwidth,angle=0
]{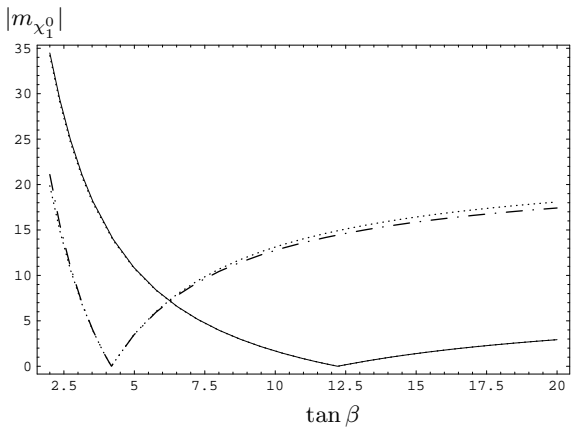}
\hspace*{4cm}{$\tan\beta$}
\caption{ The dependence of the lightest neutralino mass on $\tan\beta$ for $\lambda=0.7$ and $M_1=0.5 M_2$. 
Solid and dashed lines correspond to $M_2=\mu_{eff}=250\,\mbox{GeV}$ and $M_2=\mu_{eff}=150\,\mbox{GeV}$. 
Other notations are the same as in Fig.~1.}
\label{fig:3}       
\end{figure}

From Eq.~(\ref{17}) one can also see that the mass of the lightest neutralino decreases when $\tan\beta$ grows.
The numerical results of our analysis summarised in Figs.1--3 confirm that $|m_{\chi^0_1}|$ becomes smaller
when $\tan\beta$ raises from $3$ to $10$. However if $\tan\beta\ge \zeta=\displaystyle\frac{2 M_1 M_2 \mu_{eff}}{\tilde{M} M_Z^2}$
Eq.(\ref{17}) does not provide an appropriate description for the lightest neutralino mass. Indeed,
Eq.~(\ref{17}) implies that the mass of the lightest neutralino vanishes at large values of $\tan\beta$ while Fig.3 demonstrates
that $|m_{\chi^0_1}|$ approaches some constant non--zero value with raising of $\tan\beta$. More\, accurate\, consideration\,
of\, the approximate\, solution\, (\ref{15})\, allows\, to\, reproduce\, the\, asymptotic\, behaviour\, of\, the\, lightest\, neutralino\, 
mass\, at\, $\mu_{eff},\, M_2,\, M_1\gg M_Z$\, and\, large\, values\, of\, $\tan\beta$\, ($\tan\beta\gg \zeta$). It is given by
\begin{equation}
|m_{\chi^0_1}|\to \displaystyle\frac{\nu^2 M_Z^2}{\mu^2+\nu^2}\biggl|\frac{\tilde{M}}{M_1 M_2}\biggr|\,.
\label{18}
\end{equation}
So once again the approximate solution (\ref{15}) improves the theoretical restriction on $|m_{\chi_1^0}|$.

\section{Conclusions}
\label{sec:4} 

We have argued that the mass interval of the lightest neutralino in the MNSSM
is limited from above. The upper bound on $m_{\chi^0_1}$ has been found. In the considered 
model $|m_{\chi^0_1}|$ does not exceed $80-85\,\mbox{GeV}$ at tree level. The corresponding upper bound 
depends rather strongly on the effective $\mu$--term which is generated after the 
electroweak symmetry breaking. At large values of $|\mu_{eff}|$ the upper limit on $|m_{\chi^0_1}|$ 
goes to zero so that the mass interval of the lightest neutralino shrinks drastically.
Assuming that $|m_{\chi^0_1}|$ is considerably smaller than the masses
of the other neutralino states 
we have derived the approximate solution for the lightest neutralino mass. The obtained 
solution describes the numerical one with relatively high accuracy in most parts of
the phenomenologically allowed parameter space. Our numerical analysis and analytic considerations
show that $m_{\chi^0_1}$ decreases when $\tan\beta$ increases and the
coupling $\lambda$ decreases, respectively. 
At small values of $\lambda$ the mass of the lightest neutralino is proportional to $\lambda^2$.
The lightest neutralino mass also decreases with increasing $\mu_{eff}$, $M_1$, and $M_2$.
We have shown that at large values of the effective $\mu$--term $m_{\chi^0_1}$ is
inversely proportional to $\mu_{eff}$. In the allowed part of the parameter space the 
lightest neutralino is predominantly singlino which makes its direct observation at future colliders 
rather problematic.

\end{document}